\documentclass[letter, longauth]{aa}  

\usepackage{graphicx}
\usepackage{txfonts}
\usepackage{multirow}
\usepackage{lscape}
\usepackage{pdflscape}
\usepackage{longtable}
\usepackage{wasysym}
\usepackage{float}
\usepackage{amsmath}
\usepackage{relsize}
\usepackage{color}
\usepackage[breaklinks=true]{hyperref} 
\usepackage{float}
\usepackage{breqn}
\def\kms{km.s$^{-1}$}         
\def\ms{\hbox{m.s$^{-1}$}}         
\def\cmss{\hbox{cm.s$^{-2}$}}       
\def\kms{\hbox{km.s$^{-1}$}}       
\def\Msun{\hbox{$\mathrm{M}_{\astrosun}$}}             

\def\degr{\hbox{$^\circ$}}
\def\teff{T$_{\rm eff}$}
\def\logg{log~{\it g}}

\def\figw{\columnwidth}

\begin{document}

   \title{Spectroscopic characterisation of microlensing events\thanks{Based on observations made with ESO Telescope at the Paranal Observatory under program ID 092.C-0763(A) and 093.C-0532(A).}\fnmsep\thanks{Based on observations at Cerro Tololo Inter-American Observatory, National Optical Astronomy Observatory, which is operated by the Association of Universities for Research in Astronomy (AURA) under a cooperative agreement with the National Science Foundation.}}

   \subtitle{Towards a new interpretation of OGLE-2011-BLG-0417}
   \author{A.~Santerne\inst{\ref{IA}}\fnmsep\inst{\ref{lam}}
      \and J.-P.~Beaulieu\inst{\ref{iap}}
      \and B.~Rojas~Ayala\inst{\ref{IA}}\fnmsep\inst{\ref{santiago}}
      \and I.~Boisse\inst{\ref{lam}}
      \and E.~Schlawin\inst{\ref{Arizona}}
      \and J.-M.~Almenara\inst{\ref{ipag}}
      \and V.~Batista\inst{\ref{iap}}
      \and D.~Bennett\inst{\ref{goddard}}\fnmsep\inst{\ref{notredame}}
      \and R.~F.~D\'iaz\inst{\ref{geneva}}
      \and P.~Figueira\inst{\ref{IA}} 
      \and D.~J.~James\inst{\ref{CTIO}}\fnmsep\inst{\ref{Seattle}}
      \and T.~Herter\inst{\ref{Cornell}}
      \and J.~Lillo-Box\inst{\ref{ESO}}
      \and J.B.~Marquette\inst{\ref{iap}}
      \and C.~Ranc\inst{\ref{iap}}
      \and N.~C.~Santos\inst{\ref{IA}}\fnmsep\inst{\ref{UP}}
      \and S.~G.~Sousa\inst{\ref{IA}}
      }

   \institute{Instituto de Astrof\'isica e Ci\^{e}ncias do Espa\c co, Universidade do Porto, CAUP, Rua das Estrelas, 4150-762 Porto, Portugal\label{IA} 
	\and Aix Marseille Univ, CNRS, LAM, Laboratoire d'Astrophysique de Marseille, Marseille, France \label{lam}
	\and Institut d'Astrophysique de Paris, UMR7095 CNRS, Universit\'e Pierre \& Marie Curie, 98bis boulevard Arago, 75014 Paris, France\label{iap}
	\and Departamento de Ciencias Fisicas, Universidad Andres Bello, Fernandez Concha 700, Las Condes, Santiago, Chile\label{santiago}
	\and Steward Observatory, University of Arizona, Tucson, AZ 85721, USA\label{Arizona}
	\and UJF-Grenoble 1 / CNRS-INSU, Institut de Plan\'etologie et d'Astrophysique de Grenoble UMR 5274, Grenoble, 38041, France\label{ipag}
	\and Laboratory for Exoplanets and Stellar Astrophysics, NASA/Goddard Space Flight Center, Greenbelt, MD 20771, USA\label{goddard}
	\and Department of Physics, University of Notre Dame, 225 Nieuwland Science Hall, Notre Dame, IN 46556, USA\label{notredame}
	\and Observatoire Astronomique de l'Universit\'e de Gen\`eve, 51 chemin des Maillettes, 1290 Versoix, Switzerland\label{geneva}
	\and Cerro Tololo Inter-American Observatory, Casilla 603, La Serena, Chile\label{CTIO}
	\and Astronomy Department, University of Washington, Box 351580, U.W., Seattle, WA 98195-1580, USA\label{Seattle}
	\and Astronomy Department, Cornell University, Ithaca, NY 14853, USA\label{Cornell}
	\and European Southern Observatory (ESO), Alonso de Cordova 3107, Vitacura, Casilla 19001, Santiago de Chile, Chile\label{ESO}
	\and Departamento de F\'isica e Astronomia, Faculdade de Ci\^encias, Universidade do Porto, Rua Campo Alegre, 4169-007 Porto, Portugal\label{UP}
             }

   \date{Received 2015-11-06; Accepted 2016-10-12}

  \abstract
   {The microlensing event OGLE-2011-BLG-0417 is an exceptionally bright lens binary that was predicted to present radial velocity variation at the level of several \kms. Pioneer radial velocity follow-up observations with the UVES spectrograph at the ESO -- VLT of this system clearly ruled out the large radial velocity variation, leaving a discrepancy between the observation and the prediction. In this paper, we further characterise the microlensing system by analysing its spectral energy distribution (SED) derived using the UVES spectrum and new observations with the ARCoIRIS (CTIO) near-infrared spectrograph and the Keck adaptive optics instrument NIRC2 in the J, H, and Ks bands. We determine the mass and distance of the stars independently from the microlensing modelling. We find that the SED is compatible with a giant star in the Galactic bulge and a foreground star with a mass of 0.94$\pm$0.09\Msun\, at a distance of 1.07$\pm$0.24kpc. We find that this foreground star is likely the lens. Its parameters are not compatible with the ones previously reported in the literature (0.52$\pm$0.04\Msun\ at 0.95$\pm$0.06kpc), based on the microlensing light curve. A thoughtful re-analysis of the microlensing event is mandatory to fully understand the reason of this new discrepancy. More importantly, this paper demonstrates that spectroscopic follow-up observations of microlensing events are possible and provide independent constraints on the parameters of the lens and source stars, hence breaking some degeneracies in the analysis. UV-to-NIR low-resolution spectrographs like X-SHOOTER (ESO -- VLT) could substantially contribute to this follow-up efforts, with magnitude limits above all microlensing events detected so far.
   }


   \maketitle
%

\section{Introduction}
The gravitational microlensing is an efficient technique to detect small and cool exoplanets \citep[e.g.][]{2006Natur.439..437B} from 0.5 kpc to the Galactic bulge, in stellar populations not probed by other planet-detection techniques. However, as the other planet-detection techniques, microlensing detections are not free of false positives \citep[e.g.][]{2013ApJ...778...55H, 2016ApJ...825....8H} and degeneracy in the analysis \citep{2014ApJ...785..155B}. These effects make difficult the interpretation of the detected signals. Recent high-angular resolution follow-up observations with adaptive optics in large telescopes demonstrated that it is often possible to confirm and to refine the physical parameters of the planetary systems once the microlensing event is over. This has been achieved with the very large telescope \citep[VLT,][]{2012A&A...540A..78K}, the Keck telescope \citep{2015ApJ...808..170B}, and the Hubble space telescope \citep[HST,][]{2015ApJ...808..169B}. 

In order to test the predictive value of microlens models, \citet{2013ApJ...768..126G} proposed a new and independent route: in some cases when the lens is bright enough, one can detect the reflex motion of the lens through radial velocity observations. The system OGLE-2011-BLG-0417 \citep[hereafter OGLE-417,][]{2012ApJ...755...91S} was presented by \citet{2013ApJ...768..126G} as the best target to test this, being a relatively bright lens (V $\sim$ 18.2) with expected radial velocity variation at the level of several \kms. Following these predictions, \citet{2015A&A...582L..11B} attempted the confirmation of this system by radial velocity using the UVES spectrograph of the ESO -- VLT. The data showed no variation with a $rms$ of $\sim$ 100\ms. The reason suggested by \citet{2015A&A...582L..11B} to explain their non detection is the presence of a relatively bright star, chance-aligned with the microlensing source and lens. This star would be the one for which the radial velocities were measured. In this scenario, the lens binary is much fainter than predicted by \citet{2013ApJ...768..126G}, hence not detected in the UVES spectra. 

In this paper, we revisit the results of \citet{2012ApJ...755...91S} and \citet{2013ApJ...768..126G}, on OGLE-417 in the light of the new constraints from \citet{2015A&A...582L..11B}. We obtained new high angular-resolution imaging and near-infrared spectroscopic observations, complementing the optical UVES data already published by \citet{2015A&A...582L..11B}. These data allowed us to derive the spectral energy distribution (SED) of the few stars that are chance aligned along the line of sight with OGLE-417 (see Section \ref{data}). We analyse the SED using a method inspired from the validation of transiting exoplanets in Section \ref{SED} and present the results in Section \ref{results}. In Section \ref{comparison} we compare our results with the ones of \citet{2012ApJ...755...91S} and \citet{2013ApJ...768..126G}. Finally, in Section \ref{discut}, we draw our conclusion and discuss the results.

\section{High-resolution imaging and spectroscopic observations of OGLE-417}
\label{data}
\subsection{Keck high-resolution observations}

\begin{figure}[]
\begin{center}
\includegraphics[angle=0,width=\figw]{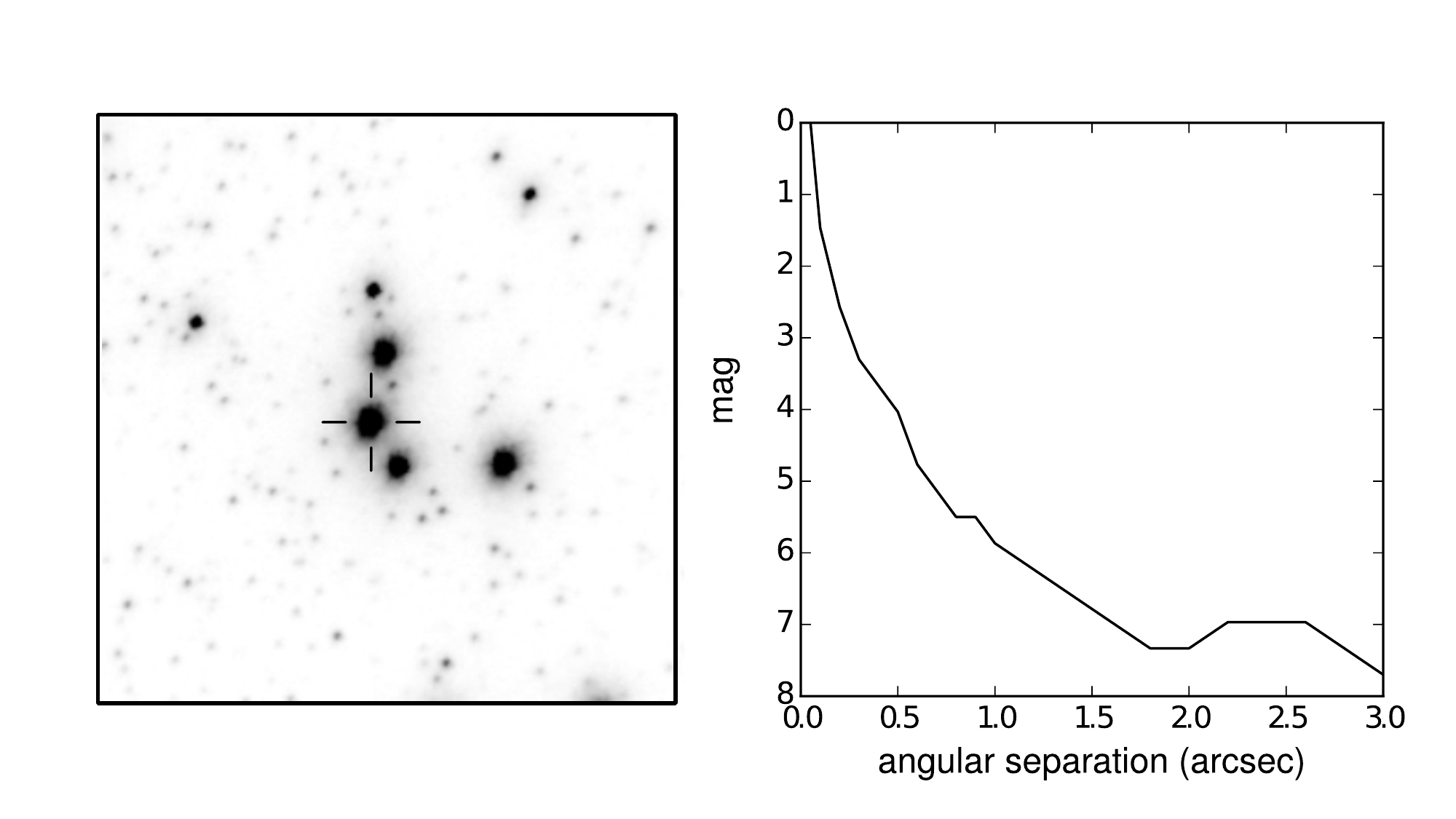}
\caption{Left: High-resolution image (5\arcsec$\times$5\arcsec) of OGLE-417 obtained by the Keck AO system in the Ks-band. North is up, East is left. The position of the microlensing event detected by OGLE is marked with the black cross. Right: 5$\sigma$ sensitivity curve from OGLE-417. Any star with a magnitude difference from OGLE-417 of less than 6 at 1\arcsec would have been significantly detected.}
\label{417AO}
\end{center}
\end{figure}

\citet{2015A&A...582L..11B} suggested that the relative bright star detected in the UVES spectra is a blend star, chanced-aligned with OGLE-417 and not the lens. To test this hypothesis, we observed the target star with the NIRC2 adaptive optics (AO) instrument at the Keck II telescope. We collected 10 Ks-band exposures with the wide camera (exposure time 10 sec), and 10 exposures with the narrow camera (exposure time 40 sec). We also collected two exposures in the J- and H-band, both with narrow and wide cameras. We processed the data following our standard procedures \citep[e.g.][]{2015ApJ...808..170B}. We display in Fig.~\ref{417AO} the Ks-band AO image at the coordinates of the microlensing event observed by OGLE. The target's PSF has a full width at half maximum (FWHM) of 130 mas. We have no clear evidence of a blend at the sub-arcsecond level. Using the method described in \citet{2014A&A...566A.103L}, we derived the 5$\sigma$ upper-limit in the presence of a blend star (see Fig~\ref{417AO}) which allow us to exclude any star within 6 magnitudes in the Ks-band at 1\arcsec\ from OGLE-417. Note that the lens and the source were expected to be still unresolved at the time of the observation. As a consequence, the blend star, if it exists, should be aligned to within a few hundreds of mas from the lens and source stars.

We measured the magnitude of the target in the three bands that we cross-matched with the out-of-magnification VVV \citep{2010NewA...15..433M} and the 2MASS \citep{2003tmc..book.....C} data using isolated stars, following the approach described in \citet{2015ApJ...808..170B}. The derived values are reported in Table \ref{mags417}. Since these magnitudes were calibrated with the VVV and 2MASS catalogs, they are in the Vega system \footnote{The flux at magnitude zero in the Vega system are of 1594~Jy, 1024~Jy, and 666.8~Jy in the J, H, and Ks band (respectively)}.

\subsection{Spectroscopic observations and data reduction}

Since the blend star suggested by \citet{2015A&A...582L..11B} was likely not detected in the Keck AO data, we decided to independently characterise the stars of the microlensing event OGLE-417. For this purpose, we used optical and near infrared (NIR) spectra of the target. The spectrum was then flux-calibrated to derive the SED.

The optical part of the spectrum was obtained with the UVES high-resolution spectrograph of the ESO -- VLT \citep{2000SPIE.4008..534D}, with the blue and red arms that cover from 0.33$\mu$m to 0.67$\mu$m. We used the spectra that were already presented in \citet{2015A&A...582L..11B}. This time however, we made use of the flux-calibrated spectra as reduced by the online pipeline \citep{2013A&A...559A..96F}. For this purpose we selected the best UVES spectrum that was obtained at low airmass, good seeing (to limit the slit loss of flux), high signal-to-noise and with no nearby contaminant significantly detected in the cross-correlation functions computed by \citet{2015A&A...582L..11B}. This spectrum was obtained on 2014-07-25. The exposure time was one hour.

The NIR part of the spectrum was obtained with the ARCoIRIS spectrograph on the Blanco telescope at CTIO \citep{2014SPIE.9147E..2HS} during commissioning nights in June 2015. ARCoIRIS is the fourth generation of the TripleSpec instrument \citep{2004SPIE.5492.1295W, 2008SPIE.7014E..0XH}, which simultaneously acquires 6 cross-dispersed orders covering $\sim$0.8 -- 2.4$\mu$m at a resolution of $\sim$3500. It has a fixed slit of 1.1$\arcsec$x28$\arcsec$ and no moving parts.

The observations were carried by placing the object at two different positions along the slit, A and B. Four exposures of 60 seconds were taken with an ABBA slit-nodding pattern. The spectra were reduced using a modified version of the Spextool reduction software \citep{2004PASP..116..362C} for ARCoIRIS. Sky-subtraction was performed by differencing A and B exposures in each paired nod. Each sky-subtracted exposure was divided by a normalized master-flat field, constructed from calibration frames taken at the beginning of the night, and wavelength calibrated using OH sky-lines. The orders 3, 4, 5, and 6 for each exposure were extracted, covering a wavelength range of 0.9-2.4$\mu$m. The 4 extracted exposures were combined and the resulting one-dimensional spectrum was telluric corrected and flux calibrated using observations of the A0V star HD 158422, obtained near in time and close in airmass to the target, with the IDL-based code \texttt{xtellcor} by \cite{2003PASP..115..389V}. 

\begin{figure}[]
\begin{center}
\includegraphics[width=\figw]{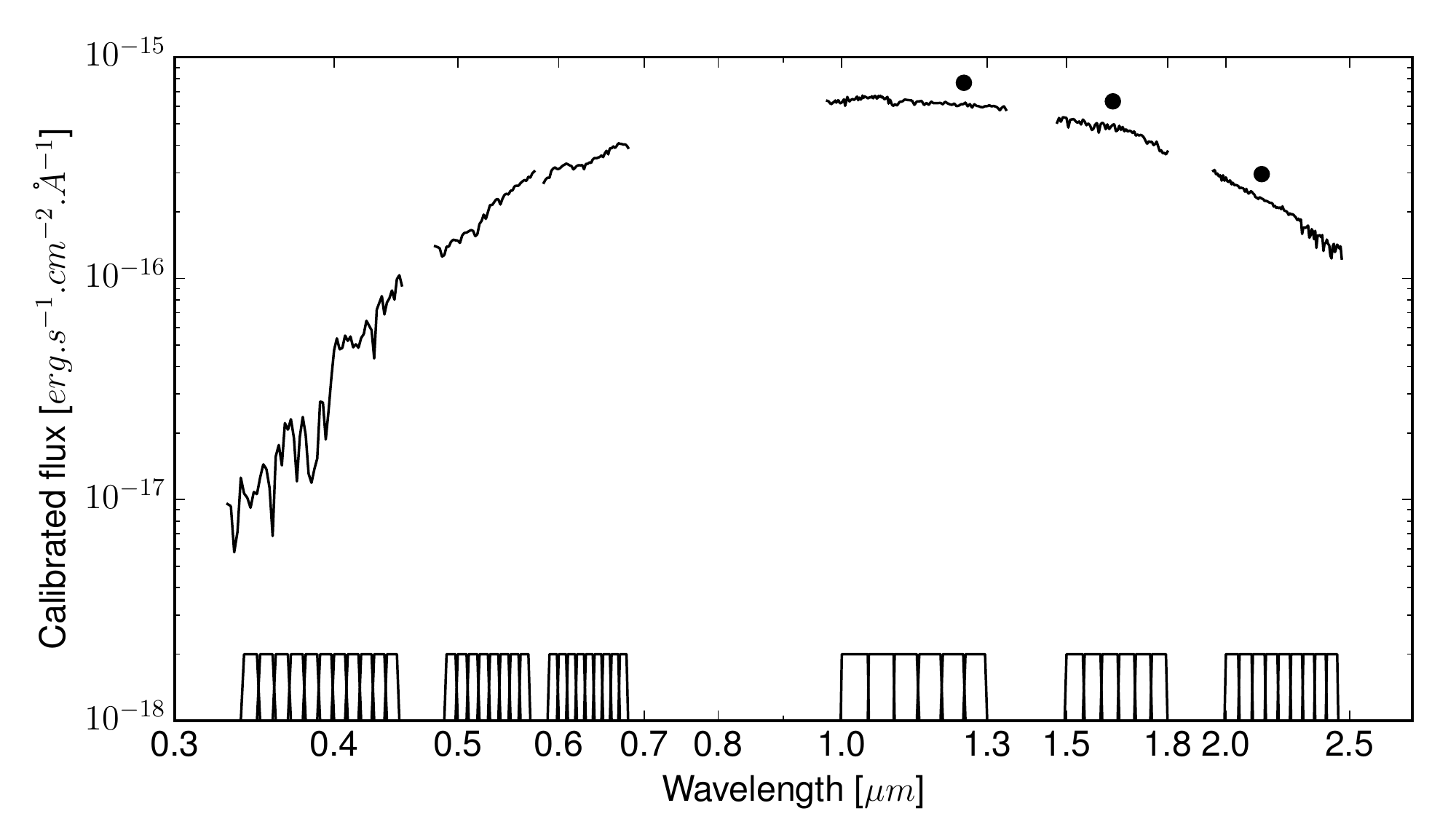}
\caption{Flux-calibrated spectrum of the target OGLE-417. The optical part was obtained with UVES at the ESO-VLT and the NIR part was obtained with ARCoIRIS at the Blanco telescope at CTIO. The black dots are the photometric magnitudes in the J, H, and Ks bands (from left to right) as measured by the Keck AO observations. The difference of flux between the NIR spectrum and the Keck AO magnitudes is due to slit loss in the spectroscopic observation. The squares in the bottom of the plot indicate the bandpasses we used to measure the SED of the target. The spectrum displayed here was binned to 20~\AA\, in the optical and 50~\AA\, in the NIR.}
\label{417Spec}
\end{center}
\end{figure}

We finally integrated the optical+NIR spectrum into small bands with a width of 100~\AA\, in the optical and 500~\AA\, in the NIR. These bands are displayed in the Fig. \ref{417Spec}. This corresponds to spectral resolutions of $\sim$50 in the optical and $\sim$40 in the NIR. The choice of these band widths was driven by two reasons. First, the target OGLE-417 is an exceptionally bright microlensing event \citep{2013ApJ...768..126G} that allows for high-resolution spectroscopic observations. Most microlensing events being much fainter, only low-resolution spectroscopy would be possible. By doing that, this technique could be applied to much fainter microlensing events. Second, we don't want to be sensitive to the presence weak stellar lines as they would add a high level of complexity into the model. We thus integrated the UVES spectrum as it was observed in low resolution. We removed the bands close or inside the UVES CCD gaps and the NIR water absorption bands. We then converted the flux into magnitude in the AB system\footnote{The flux at magnitude zero is of 3631~Jy for all bands} that we list in the Table \ref{mags417}. We propagated the errors from the data reduction and flux calibration to the final magnitudes. We finally added 20mmag of possible instrumental systematics to the errors.

\section{SED Analysis}
\label{SED}

We used the \texttt{PASTIS} software \citep{2014MNRAS.441..983D, 2015MNRAS.451.2337S} to model the SED. \texttt{PASTIS} is designed to validate transiting exoplanets by estimating their probability against false-positive scenarios \citep[such as blended eclipsing binaries, see e.g.][]{2014A&A...571A..37S}. It uses the SED to constrain the relative magnitude and color of potential blends. The modelling of the SED into the \texttt{PASTIS} software is fully described in \citet{2014MNRAS.441..983D}. It has already been used in e.g. \citet{2014MNRAS.444.2783M}, \citet{2014A&A...571A..37S}, \citet{2015A&A...582A..33A}, and \citet{2015arXiv150602471D}. For the sake of clarity, we present below the modelling and analysis of the SED. 

The SED was modelled with the BT-SETTL stellar atmosphere models of \citet{2012IAUS..282..235A} that we integrated into the same bandpasses as for the optical+NIR spectrum. We used the Dartmouth stellar evolution tracks from \citet{2008ApJS..178...89D} to determine the stellar atmospheric parameters from the fundamental parameters. For the interstellar extinction, we used the model of \citet{2005AJ....130..659A} that we computed for the line of sight of OGLE-417. We added the interstellar extinction to the BT-SETTL models following the law of \citet{1999PASP..111...63F}. 

We modelled the SED with a giant star in the Galactic bulge that is assumed to be the source of the microlensing event, and a foreground star that is chance-aligned within 1\arcsec and thus, fully contributes to the SED\footnote{The slit size of the UVES and ARCoIRIS observations were of 1\arcsec\ and 1.1\arcsec, respectively.}. We analysed the data through the Markov Chain Monte Carlo (MCMC) algorithm of the \texttt{PASTIS} software which is described in \citet{2014MNRAS.441..983D}. We used a Gaussian prior for the source star with typical parameters for a giant star in the bulge, i.e. a \teff\, of 4660 $\pm$ 250 K and a \logg\, of 2.5 $\pm$ 0.5 dex, as in \citet{2012ApJ...755...91S}. We assumed a prior for the mass of the foreground star following the initial mass function of \citet{2001MNRAS.322..231K}. For the other parameters, we used non-informative priors. The interstellar extinction of the foreground star is fixed to the value from \citet{2005AJ....130..659A} which depends on the distance. For the source however, we let it as a free parameter in the analysis.

We cut the SED in three chunks: one for the optical magnitudes (derived from UVES), and two for the NIR magnitudes (one from the ARCoIRIS data and another one from the Keck AO magnitudes). We fit in the MCMC procedure a possible slit loss for the two sets of SED derived with the UVES and ARCoIRIS data. These two slit loss factors are constrained thanks to the Keck AO data. We also include an extra source of white noise for the magnitudes of the three sets of SED. These parameters were let free in the analysis. The entire list of parameters and their respective priors are reported in Table \ref{tableparams}.

We ran 100 MCMC chains of $10^{6}$ iterations randomly drawn from the joint prior distribution. All chains converged towards the same solution which is assumed to be the global maximum of the posterior. After removing for the burn-in phase, the chains were thinned and merged together. The final posterior distribution has more than 15000 independent samples. The median values and their 68.3\% uncertainties are reported in Table \ref{tableparams}.

We note that this method is different from the one presented in \citet{1998A&A...338...56M} which requires observations at different epochs of the magnification. This method is similar to the one described in \citet{2013A&A...555A..16T} except that we used both flux-calibrated spectra and high spatial resolution imaging and not broadband aperture photometry.

\section{Results}
\label{results}

\begin{figure}[h]
\begin{center}
\includegraphics[width=\figw]{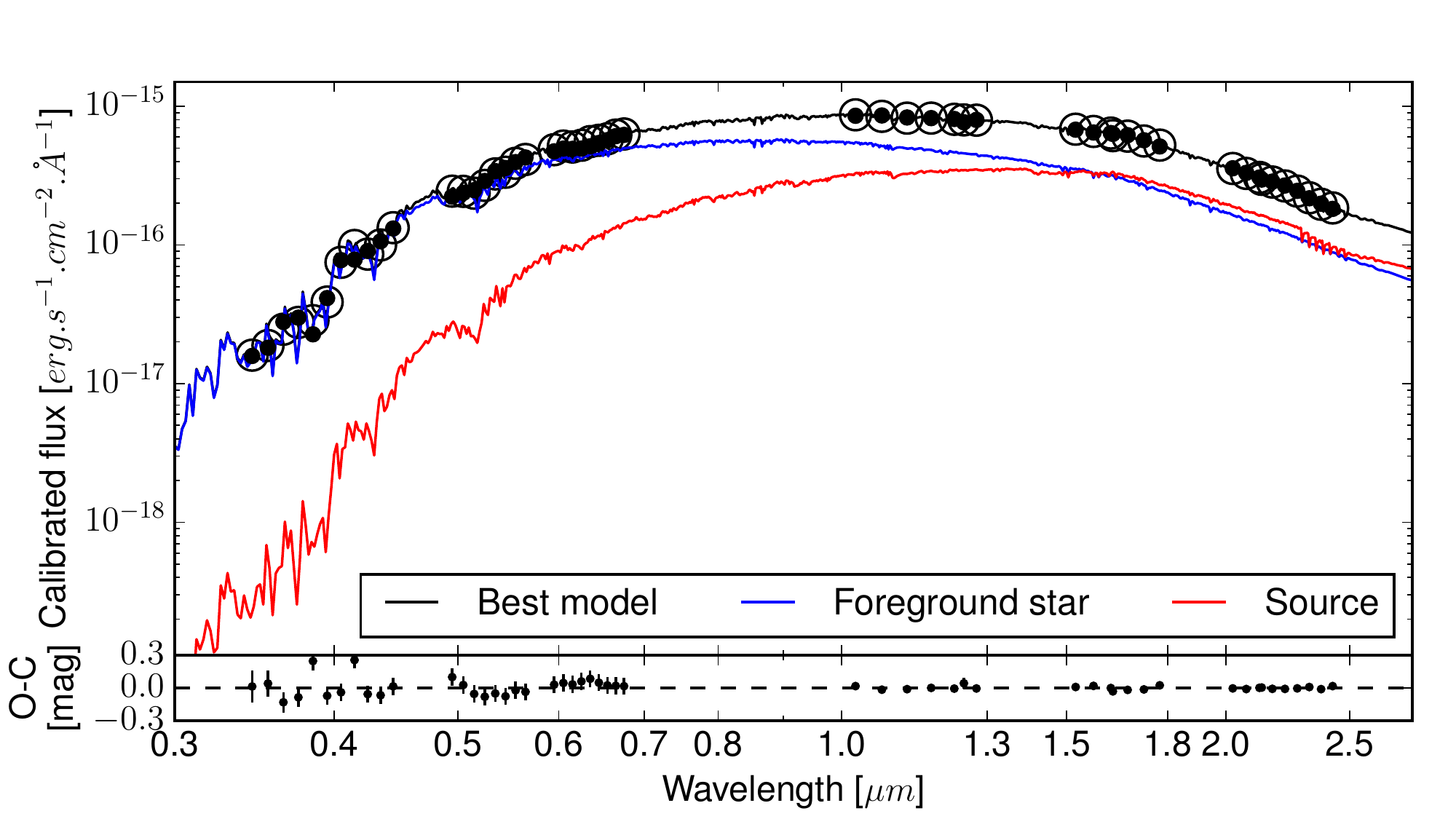}
\caption{Spectral energy distribution of OGLE-417 together with the best-fit model. The black dots are the measured SED while the open circles are the integrated model in the corresponding bandpasses. The black line is the best model that fit the spectroscopic data and the red and blue lines are the individual model of the source and foreground star (respectively). The bottom panel shows the residuals to the best fit.}
\label{417SED}
\end{center}
\end{figure}

We find that the SED of OGLE-417 is well reproduced with a scenario of a giant source star located at 8.77$^{_{+0.90}}_{^{-1.40}}$~kpc, and a foreground star of about 0.94 \Msun\, at $\sim$1.1kpc. The measured value of the interstellar extinction for the source star, E(B-V) = 1.21$\pm$0.16, which corresponds to A(I) = 1.90$\pm$0.25 according to \citet{1999PASP..111...63F}, is in very good agreement with the value used by \citet{2012ApJ...755...91S} of A(I)$\sim$2.0. The best fitted model is displayed in Fig. \ref{417SED}, together with the SED of the individual stars. 

From the posterior samples, we also derived the apparent magnitudes of both the foreground and source stars in different bandpasses that we report in Table \ref{mags}. What we call the foreground star in our model has a V magnitude of $\sim$17.7 and is bluer than the source which has a V magnitude of $\sim$19.4. As a consequence, the bluest star with a deep line profile detected by \citet[see their Fig. 2]{2015A&A...582L..11B} is the foreground star and the reddest star with a shallow line profile is the microlensing source. This is the opposite of what they reported, which was based on the incorrect color information published by \citet{2012ApJ...755...91S} and \citet{2013ApJ...768..126G}. This does not affect the result of \citet{2015A&A...582L..11B}, only the sign of their radial velocity curve.

\begin{figure}[]
\begin{center}
\includegraphics[width=\figw]{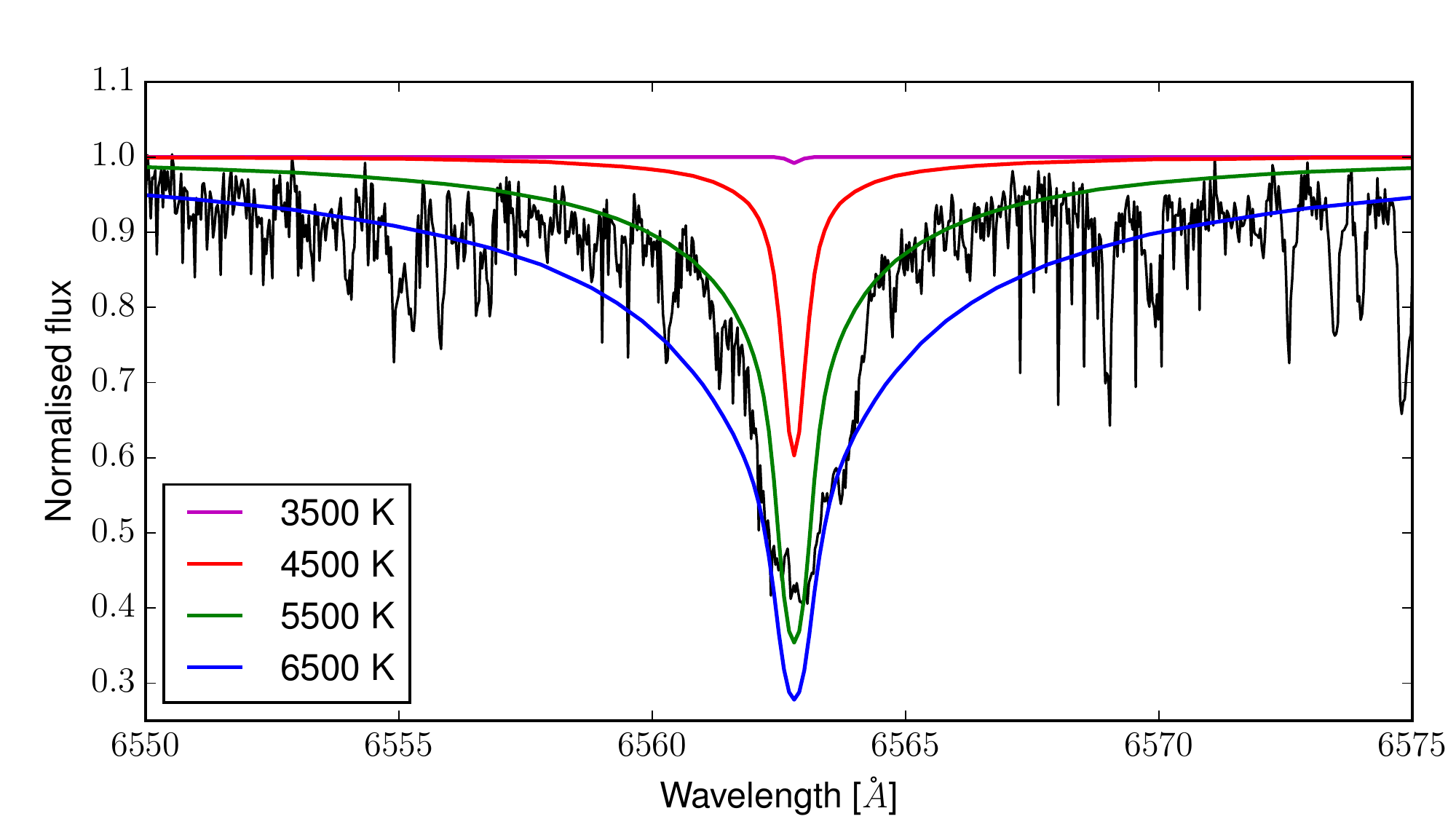}\\
\includegraphics[width=\figw]{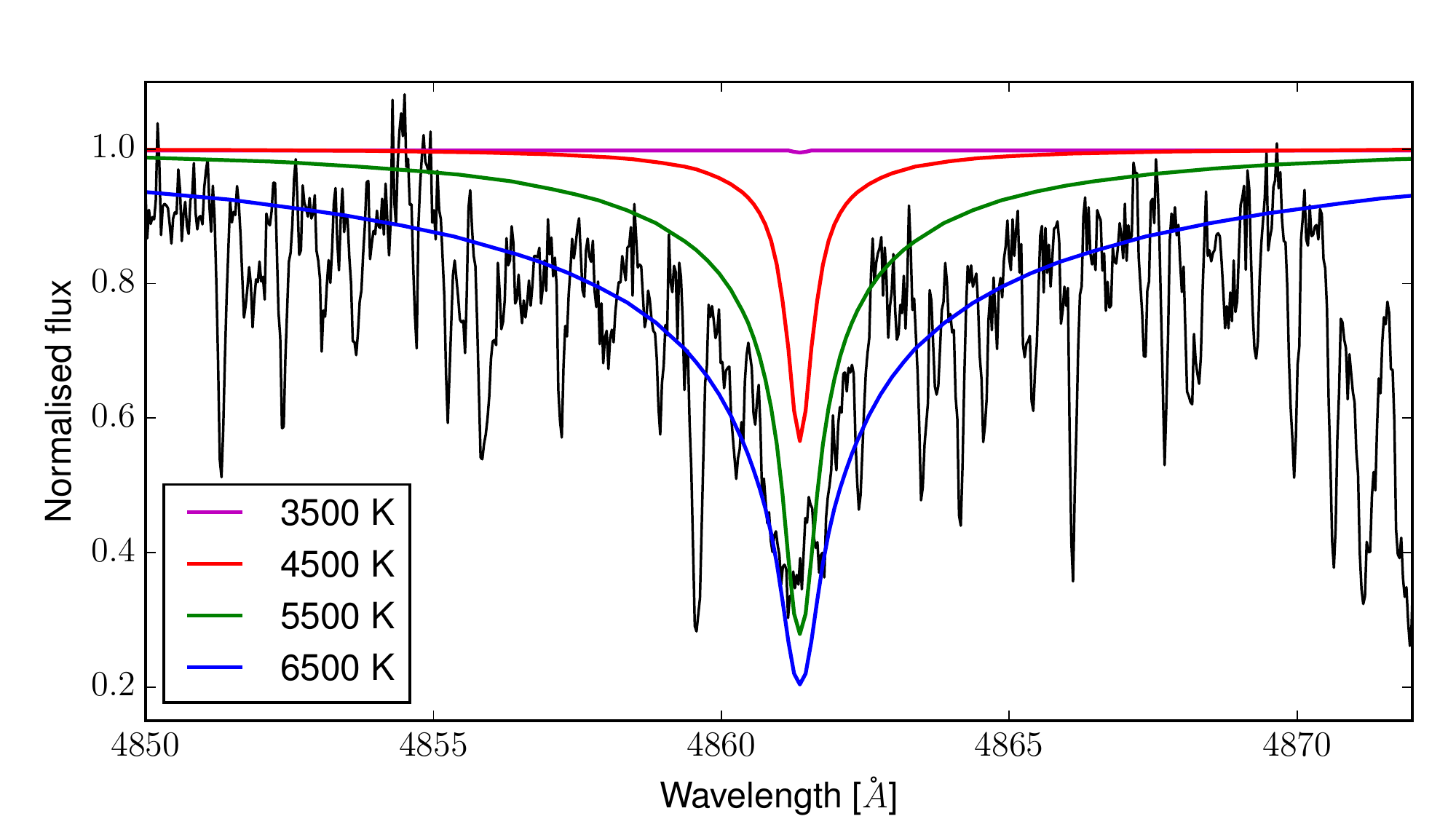}
\caption{High-resolution, co-added and normalised UVES spectrum of OGLE-417 (in black) of the temperature-sensitive H$_\alpha$ (top) and H$_\beta$ (bottom) lines. Four models from \citet{1993ASPC...44...87K} of the Balmer line profile for \teff\, of 3500~K (magenta), 4500~K (red), 5500~K (green), and 6500~K (blue) are also displayed. The red wing of the H$_\alpha$ line has an asymmetric shape due to the presence of the source star. The contrast between the foreground and source star is higher in the blue (see Fig. \ref{417SED}). Thus, the contribution from the source is lower for the H$_\beta$ line.}
\label{417HR}
\end{center}
\end{figure}

Using the Dartmouth evolution tracks, we derived from the posterior samples that the foreground star has a \teff~=~5430~$\pm$~140~K, a \logg~=~4.46$^{_{+0.08}}_{^{-0.12}}$~\cmss, and the metallicity reported in Table \ref{tableparams}. This corresponds to a mid-G dwarf. As a sanity check of our results, we co-added the high-resolution UVES spectra, after correcting for the barycentric Earth radial velocity and from the systemic radial velocity of the foreground star, as measured by \citet{2015A&A...582L..11B}. We then normalised it in the vicinity of the \teff-sensitive Balmer H$_{\alpha}$ and H$_{\beta}$ lines. This co-added spectrum is displayed in Fig. \ref{417HR}. At these wavelengths, the foreground star is the brightest star, and should dominate the spectrum. However, the contribution from the source star which is red-shifted by about 42~\kms\ \citep[hence of about 1~\AA,][]{2015A&A...582L..11B} is clearly visible in the red part of the H$_{\alpha}$ spectrum. It makes the analysis of the high-resolution spectrum with classical spectroscopic techniques \citep[e.g.][and references therein]{2014arXiv1407.5817S} not reliable. Given that the source is a cool, giant star, the shape of the H$_{\alpha}$ blue wing of the foreground star is not expected to be substantially affected. Given that the flux ratio between both stars is higher in the blue, the source contribution in the H$_{\beta}$ line is expected to be significantly lower than the one of H$_{\alpha}$.

We compared this high-resolution spectrum with theoretical LTE models from \citet{1993ASPC...44...87K}\footnote{http://kurucz.harvard.edu/grids/gridp01/bp01k2.datcd} for \teff\ ranging from 3500~K to 6500~K and a fixed logg of 4.5~\cmss\ (see Fig. \ref{417HR}). These basic models are supposed to reproduce correctly the relative intensity and the wings of the Balmer lines \citep{2008AN....329..573A}. As displayed in Fig. \ref{417HR}, the blue wing of H$_{\alpha}$ line and the H$_{\beta}$ line of the foreground star correspond to the one of a mid-G dwarf, with a \teff\ of about 5500~K. A cooler or hotter star would have produced a weaker or stronger line (respectively). This fully supports the results of our SED analysis.

As a second sanity check for our results, we compared the magnitude of the foreground + source stars in the I band as predicted by our SED model with the one observed by \citet{2012ApJ...755...91S}. This constraint was not used in our SED analysis. \citet{2012ApJ...755...91S} reported an out-of-magnification apparent I-band magnitude of 15.74\footnote{http://ogle.astrouw.edu.pl/ogle4/ews/2011/blg-0417.html}. No associated error nor the reference system (Vega or AB) is provided with this magnitude. In the Vega system, our model predicts that the apparent I-band magnitude of the foreground+source stars is of 15.16~$\pm$~0.06 (see Table \ref{mags}). In the AB system \citep{1974ApJS...27...21O}, the apparent I-band magnitude of this system is of 15.67~$\pm$~0.06. As a consequence, our model is in perfect agreement with the magnitude measured by OGLE in the I band, assuming it is provided in the AB system.

\section{Is the foreground star the lens or a blend ?}
\label{comparison}

\begin{figure}[]
\begin{center}
\includegraphics[width=\figw]{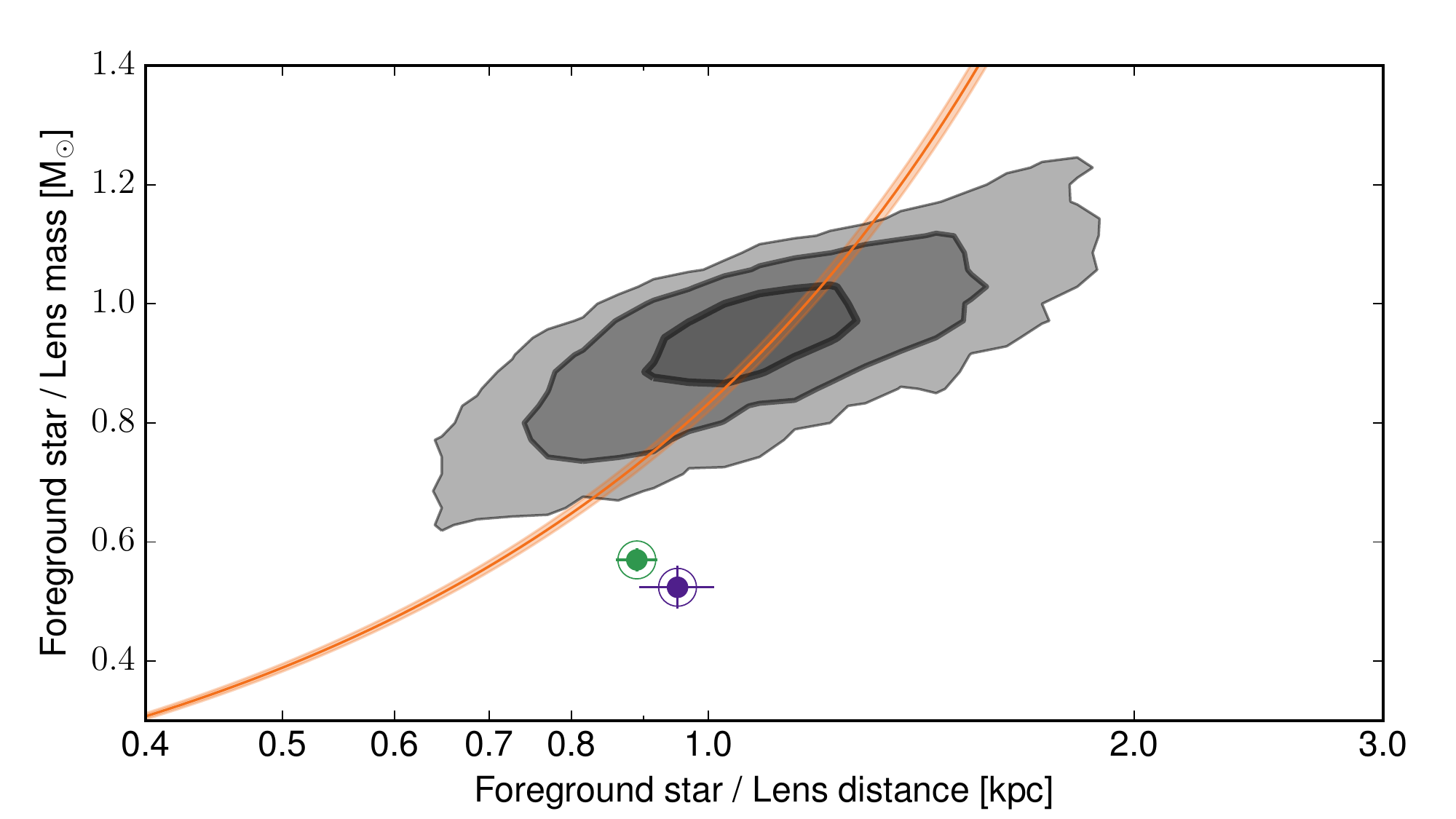}
\caption{Posterior distribution of the foreground star mass as a function of its distance. The grey regions correspond to the 68.3\%, 95.5\%, and 99.7\% (from dark to light greys) of confidence intervals. The green and purple marks are the positions of the lens primary as reported by \citet{2012ApJ...755...91S} and \citet{2013ApJ...768..126G}, respectively. The orange line shows the lens total mass vs distance relation for  $\Theta_E=2.44 \pm 0.02$ mas and $d_{S} = 8.2$ kpc.}
\label{417Blend}
\end{center}
\end{figure}

In our SED analysis, we find that the foreground star, the one for which radial velocity were measured by \citet{2015A&A...582L..11B}, is a mid-G dwarf at about 1~kpc. In this section we discuss the nature of this foreground star in the context of the microlensing event. In Fig. \ref{417Blend} we show the posterior distribution of the foreground star together with the position of the lens primary as predicted by \citet{2012ApJ...755...91S} and \citet{2013ApJ...768..126G}. The mass of the foreground star we derived is significantly different (by more than 4$\sigma$) from the masses of the lens primary as reported by these authors. This could be explained by two main reasons: (1) the foreground star is a fourth, blend star, chance-aligned with the source and the lens binary, as suggested by \citet{2015A&A...582L..11B}, (2) the foreground star is the lens and its parameters were incorrectly determined in both \citet{2012ApJ...755...91S} and \citet{2013ApJ...768..126G}. 

The first scenario could explain the absence of radial velocity variation reported by \citet{2015A&A...582L..11B}, as the lens would be too faint to be detected in the UVES data. To test this scenario, we analysed the SED using a more complex model composed of a source, a foreground star, and a faint binary system that would correspond to the lens. There is no evidence in the data for this faint binary system, either bound or not with the foreground star, so we can not rule it out.

Even if the density of stars is very high towards the galactic center, it is quite unlikely, \textit{a priori}, to have a system with a source star, a binary lens, and a blend star chance aligned within the contraints of the Keck AO observations (see Fig. \ref{417AO}). In the Ks band, the blend has a magnitude of 13.87 $\pm$ 0.22. We collected all the stars listed in the VVV DR1 \citep{2010NewA...15..433M} that are within 1\degr\ of OGLE-417. We assume here an homogeneous distribution of the stars within this 1\degr\, of radius. We estimated that the \textit{a priori} probability to have a blend star in the range Ks~$\in$~[13.21 ; 14.53] (hence within 3$\sigma$ of the value derived by the SED analysis and within the 5$\sigma$ sensitivity curve shown in Fig. \ref{417AO}), is at the level of 110ppm. Given that there is no evidence in the SED for a M dwarf binary at $\sim$~900~pc and that the presence of a chance-aligned blend star is \textit{a priori} unlikely, we reject this scenario.

The second scenario is apparently not compatible with the absence of radial velocity variation observed by \citet{2015A&A...582L..11B}. Indeed, if the foreground star is the lens, it would have exhibit significant radial velocity variations, unless the system parameters derived by the analysis of the microlensing light curve are incorrect. However, it is interesting to note that \citet{2012ApJ...755...91S} reported an Einstein radius of the microlensing event to be $\Theta_E=2.44 \pm 0.02$ mas. It is related to the lens physical parameters following the mass-distance relation \citep{2000ApJ...542..785G}:
\begin{equation}
M_{T} = \Theta_E^2 \frac{c^{2}}{4G} d_{L}\left(1-\frac{d_{L}}{d_{S}}\right)^{-1}\ ,
\label{MRrelation}
\end{equation}
where $M_{T}$ is the total mass of the lens system and $d_{L}$ and $d_{S}$ are the distances of the lens and the source (respectively). We find that our constraints on the foreground star are perfectly compatible with this mass-distance relation for $d_{S} = 8.2$ kpc and assuming a negligible mass for the lens companion (see Fig \ref{417Blend}). We therefore conclude that this foreground star is likely the lens, and that the system parameters derived by \citet{2012ApJ...755...91S} and \citet{2013ApJ...768..126G} are incorrect. In this scenario, the lens companion needs to have a low mass, or the orbit needs a high inclination to explain the absence of significant radial velocities variation found by \citet{2015A&A...582L..11B}. 

A third scenario might also be drawn, in which the foreground star is bound with a pair of faint M dwarfs. In this scenario, the lens would be a hierarchical triple system. It is however expected that such triple system would have significantly affected the microlensing light curve, leading to incorrect parameters as derived by \citet{2012ApJ...755...91S} and \citet{2013ApJ...768..126G} in their binary model.

If the lens system parameters reported by \citet{2012ApJ...755...91S} and \citet{2013ApJ...768..126G} are incorrect, it is likely that the value for the Einstein radius is also incorrect. This would limit the above comparison. Using our constraints on the mass and distance for the foreground star, assumed here to be the lens, and considering it has a companion with a mass ratio ranging from zero to one, we can use the equation \ref{MRrelation} to constrain $\Theta_{E}$. In this case, we find that $\Theta_{E} \in [1.9 ; 3.6]$ mas to be compatible within 3$\sigma$ of our spectroscopic results.

\section{Discussion and conclusions}
\label{discut}

In this paper, we used an optical+NIR low-resolution (R $\sim$ 40 -- 50) flux-calibrated spectrum obtained with the UVES (ESO -- VLT) and ARCoIRIS (Blanco telescope -- CTIO) spectrographs to analyse the SED of the microlensing event OGLE-417. We estimate the slit loss of these spectra using the uncontaminated magnitude of the target in the J, H, and Ks bands measured with the Keck AO facility NIRC2. This also allowed us to constrain the absence of additional stars in the immediate vicinity of the target.

We find that the SED is compatible with a scenario of a source giant as predicted by \citet{2012ApJ...755...91S} and a foreground star of $\sim$~0.94~\Msun\, located at $\sim$ 1.1~kpc. This foreground star is the one observed in radial velocity by \citet{2015A&A...582L..11B} together with the source star. Its parameters are fully compatible with the mass and distance of the lens assuming an Einstein radius of $\Theta_E=2.44 \pm 0.02$ mas \citep{2012ApJ...755...91S}, a source at 8.2 kpc, and a very-low companion mass. This is however not compatible with the lens parameters derived by \citet{2012ApJ...755...91S} and \citet{2013ApJ...768..126G}. A reanalysis of this microlensing light curve is mandatory to fully understand the discrepancies between its modelling and the spectroscopic results \citep[and this work]{2015A&A...582L..11B}. This is however out of the scope of this paper.

The information provided by \citet{2012ApJ...755...91S} and \citet{2013ApJ...768..126G} suggests that the modeling effort was not sufficient to find all the possible solutions. There is no indication of an effort to probe for multiple solutions in the orbital and microlensing parallax parameter space. Some degeneracy between the orbital motion and parallax effects is to be expected, but there is no discussion of this. There is also no indication of the exploration of  alternate models, such as triple lens models or binary source models. Finally, some of the reported error bars are suspiciously small, such as the error bar on the line-of-sight separation at 2\% of the Einstein radius. This is 30 times smaller than the uncertainty reported by \citet{2011ApJ...738...87S}, and it suggests that the MCMC used by \citet{2012ApJ...755...91S} is not well mixed and weakly account for correlated parameter space.

This paper also demonstrates that the spectroscopic characterisation of microlensing events is possible by fitting spectral energy distributions to a low-resolution, flux-calibrated spectrum. This technique could support the characterisation of microlensing events by providing independent constraints on the source, lens, and possible blend star properties, and thus help to break some degeneracies in the analyses. For that, it is however important to have a spectrum covering the optical and NIR wavelengths. In case of crowded fields, AO observations are also important to calibrate the slit loss and the absolute flux of the target. Low-resolution UV-to-NIR spectrographs like X-SHOOTER \citep{2011A&A...536A.105V} at the ESO -- VLT, with a magnitude limit down to about 21, would allow one to use this method to characterise all of the microlensing events.

\begin{acknowledgements}
We thank the anonymous referee for their fruitful comments. Some of the data presented herein were obtained at the W.M. Keck Observatory, which is operated as a scientific partnership among the California Institute of Technology, the University of California and the National Aeronautics and Space Administration. The Observatory was made possible by the generous financial support of the W.M. Keck Foundation. The authors wish to recognize and acknowledge the very significant cultural role and reverence that the summit of Mauna Kea has always had within the indigenous Hawaiian community. 
Astronomy Research using the Cornell Infra Red Imaging Spectrograph (ARCoIRIS) was made possible through supplemental funding from the National Science Foundation to the NOAO under the "Renewing Small Telescopes for Astronomical Research (ReSTAR)" Phase 1 program (US Federal Award ID: 0936648).
The Porto group acknowledges the support from Funda\c{c}\~ao para a Ci\^encia e a Tecnologia (FCT, Portugal) in the form of grants and Investigador FCT contracts of reference PTDC/FIS-AST/1526/2014 (POCI-01-0145-FEDER-016886), SFRH/BPD/87776/2012, IF/00169/2012, IF/01037/2013, IF/00028/2014, and UID/FIS/04434/2013 (POCI-01-0145-FEDER-007672), as well as POPH/FSE (EC) by FEDER funding through the program ``Programa Operacional de Factores de Competitividade - COMPETE''.  AS is supported by the European Union under a Marie Curie Intra-European Fellowship for Career Development with reference FP7-PEOPLE-2013-IEF, number 627202. JMA acknowledges funding from the European Research Council under the ERC Grant Agreement n. 337591-ExTrA. V. B. was supported by the CNES and the DIM ACAV, Region Ile de France. V.B., J.P.B., J.B.M. acknowledge the support of PERSU Sorbonne Universit\'e and the Programme National de Plan\'etologie. B.R-A acknowledges the support from CONICYT PAI/CONCURSO NACIONAL INSERCI\'ON EN LA ACADEMIA, CONVOCATORIA 2015 79150050.
\end{acknowledgements}

\Online

\begin{table*}[h]
\caption{Priors and posteriors defined in the \texttt{PASTIS} analyses: $\mathcal{U}(a,b)$ represents a Uniform prior between $a$ and $b$; $\mathcal{N}(\mu,\sigma^{2})$ represents a Normal distribution with a mean of $\mu$ and a width of $\sigma^{2}$; $\mathcal{P}(\alpha; x_{min}; x_{max})$ represents a Power Law distribution with an exponent $\alpha$ computed between $x_{min}$ and $x_{max}$ ; $\mathcal{P}_{2}(\alpha_{1}; \alpha_{2}; x_{0}; x_{min}; x_{max})$ represents a double Power Law distribution with an exponent $\alpha_{1}$ computed between $x_{min}$ and $x_{0}$ and an exponent $\alpha_{2}$ computed between $x_{0}$ and $x_{max}$; and finally $\mathcal{S}(a,b)$ represents a Sine distribution between $a$ and $b$.}
\begin{center}
\begin{tabular}{lcc}
\hline
\hline
Parameter & Prior & Posterior \\
\hline
\multicolumn{3}{l}{\textit{Source star}}\\
&& \\
Effective temperature \teff\, [K] & $\mathcal{N}(4660; 250)$ & 4585 $\pm$ 140 \\
Surface gravity \logg\, [g.cm$^{-2}$] & $\mathcal{N}(2.5; 0.5)$ & 2.54 $\pm$ 0.16 \\
Iron abondance [Fe/H]$_{S}$ [dex] & $\mathcal{U}(-2.5; 0.5)$ & 0.10 $\pm$ 0.34 \\
Distance $D_{S}$ [pc] & $\mathcal{P}(2; 6000; 10000)$ & 8770$^{_{+900}}_{^{-1400}}$\\
Interstellar extinction E(B - V) [mag] & $\mathcal{U}(1; 5)$ & 1.21 $\pm$ 0.16\\
\hline
\multicolumn{3}{l}{\textit{Foreground / Lens star}}\\
&& \\
Initial mass $M_{L}$ [\Msun] & $\mathcal{P}_{2}(-1.3; -2.3; 0.5; 0.1; 20)$ & 0.94 $\pm$ 0.09 \\
Iron abondance [Fe/H]$_{L}$ [dex] & $\mathcal{U}(-2.5; 0.5)$ & 0.17$^{_{+0.19}}_{^{-0.28}}$\\
Age $\tau_{L}$ [Gyr] & $\mathcal{U}(0.1; 13.7)$ & 7.3 $\pm$ 4.5 \\
Distance $D_{L}$ [pc] & $\mathcal{P}(2; 10; 6000)$ & 1070 $\pm$ 240 \\
\hline
\multicolumn{3}{l}{\textit{Others}}\\
&& \\
Optical extra white noise $\sigma_{OPT}$ [mag] & $\mathcal{U}(0; 1)$ & 0.09 $\pm$ 0.02\\
NIR extra white noise $\sigma_{NIR}$ [mag] & $\mathcal{U}(0; 1)$ & 0.005$^{_{+0.005}}_{^{-0.003}}$\\
Keck AO extra white noise $\sigma_{keck}$ [mag] & $\mathcal{U}(0; 1)$ & 0.06$^{_{+0.12}}_{^{-0.04}}$\\
Fraction of UVES flux $f_{UVES}$ & $\mathcal{U}(0; 1)$ & 0.66 $\pm$ 0.06\\
Fraction of ARCoIRIS flux $f_{ARCoIRIS}$ & $\mathcal{U}(0; 1)$ & 0.76 $\pm$ 0.04\\
\hline
\hline
\end{tabular}
\end{center}
\label{tableparams}
\end{table*}%

\begin{table}[]
\caption{Apparent magnitudes of the foreground and source stars derived from the posterior distribution.}
\begin{center}
\begin{tabular}{lccc}
\hline
\hline
Band & Foreground & Source & Foreground+Source\\
\hline
B & 19.27 $\pm$ 0.17 & 21.72 $\pm$ 0.72 & 19.16 $\pm$ 0.10\\
V & 17.70 $\pm$ 0.20 & 19.40 $\pm$ 0.58 & 17.49 $\pm$ 0.09\\
R & 16.53 $\pm$ 0.20 & 17.64 $\pm$ 0.44 & 16.19 $\pm$ 0.07\\
I & 15.64 $\pm$ 0.20 & 16.30 $\pm$ 0.34 & 15.16 $\pm$ 0.06\\
J & 14.64 $\pm$ 0.21 & 14.76 $\pm$ 0.24 & 13.94 $\pm$ 0.05\\
H & 14.06 $\pm$ 0.21 & 13.86 $\pm$ 0.20 & 13.20 $\pm$ 0.05\\
Ks & 13.87 $\pm$ 0.22 & 13.55 $\pm$ 0.17 & 12.94 $\pm$ 0.05\\
\hline
\hline
\end{tabular}
\tablefoot{All these magnitudes are provided in the Vega system assuming a zero-magnitude flux of 4024~Jy, 3563~Jy, 2815~Jy, 2283~Jy, 1594~Jy, 1024~Jy, and 667~Jy for the B, V, R, I, J, H, and Ks band (respectively).}
\end{center}
\label{mags}
\end{table}%

\begin{table}[h]
\caption{Magnitudes of the target OGLE-2011-BLG-0417. The spectral bands J, H, and Ks are referenced in the Vega system, while the other ones are referenced in the AB system, i.e. the zero-magnitude corresponds to a flux of 3631 Jy. The spectral domains of the custom bands are expressed in Angstrom. The lines indicate the limits between the three sets of SED.}
\begin{center}
\begin{tabular}{ccc}
\hline
\hline
Spectral band & Magnitude & Error\\
\hline
J & 13.98 & 0.05\\
H & 13.16 & 0.03\\
Ks & 12.94 & 0.03\\
\hline
3400 -- 3500  &  22.372  &  0.129\\
3500 -- 3600  &  22.163  &  0.091\\
3600 -- 3700  &  21.637  &  0.053\\
3700 -- 3800  &  21.501  &  0.043\\
3800 -- 3900  &  21.744  &  0.045\\
3900 -- 4000  &  21.035  &  0.030\\
4000 -- 4100  &  20.296  &  0.024\\
4100 -- 4200  &  20.235  &  0.023\\
4200 -- 4300  &  20.027  &  0.023\\
4300 -- 4400  &  19.801  &  0.023\\
4400 -- 4500  &  19.522  &  0.023\\
4900 -- 5000  &  18.717  &  0.021\\
5000 -- 5100  &  18.605  &  0.020\\
5100 -- 5200  &  18.513  &  0.020\\
5200 -- 5300  &  18.306  &  0.020\\
5300 -- 5400  &  18.087  &  0.020\\
5400 -- 5500  &  17.991  &  0.020\\
5500 -- 5600  &  17.845  &  0.020\\
5600 -- 5700  &  17.723  &  0.020\\
5900 -- 6000  &  17.500  &  0.020\\
6000 -- 6100  &  17.414  &  0.020\\
6100 -- 6200  &  17.383  &  0.020\\
6200 -- 6300  &  17.343  &  0.020\\
6300 -- 6400  &  17.267  &  0.020\\
6400 -- 6500  &  17.174  &  0.020\\
6500 -- 6600  &  17.089  &  0.020\\
6600 -- 6700  &  16.985  &  0.020\\
6700 -- 6800  &  16.929  &  0.020\\
\hline
10000 -- 10500    &   15.511   &    0.021\\
10500 -- 11000    &   15.402   &    0.021\\
11000 -- 11500    &   15.340   &    0.021\\
11500 -- 12000    &   15.257   &    0.020\\
12000 -- 12500    &   15.182   &    0.020\\
12500 -- 13000    &   15.112   &    0.021\\
15000 -- 15500    &   14.899   &    0.020\\
15500 -- 16000    &   14.880   &    0.020\\
16000 -- 16500    &   14.822   &    0.020\\
16500 -- 17000    &   14.794   &    0.020\\
17000 -- 17500    &   14.822   &    0.020\\
17500 -- 18000    &   14.872   &    0.021\\
20000 -- 20500    &   14.979   &    0.021\\
20500 -- 21000    &   15.009   &    0.021\\
21000 -- 21500    &   15.041   &    0.021\\
21500 -- 22000    &   15.064   &    0.021\\
22000 -- 22500    &   15.085   &    0.021\\
22500 -- 23000    &   15.134   &    0.021\\
23000 -- 23500    &   15.220   &    0.022\\
23500 -- 24000    &   15.271   &    0.023\\
24000 -- 24500    &   15.320   &    0.032\\
\hline
\hline
\end{tabular}
\end{center}
\label{mags417}
\end{table}%

\end{document}